# Correlated Structural and Optical Characterization during Van der Waals Epitaxy of PbI$_2$ on graphene

C.P. Sonny Tsotezem[1*], E. M. Staicu Casagrande[1], A. Momeni[1,2], A. Ouvrard[1], A. Ouerghi[3], M. Rosmus[1], A. Antezak[1], F. Fortuna[1], A. F. Santander-Syro[1], E. Frantzeskakis[1], A.M. Lucero Manzano[4], E.D. Cantero[4], E.A. Sánchez[4], H. Khemliche[1*]

[1] Institut des Sciences moleculaires d'Orsay, Université Paris-Saclay, CNRS, 91405, Orsay, France

[2] CY Cergy Paris Université, 95000 Cergy, France

[3] Centre de Nanosciences et de Nanotechnologies, 91120, PalaiseauISMO, Université Paris-Saclay, CNRS, 91405, Orsay, France

[4] Instituto de Nanociencia y Nanotecnología - Nodo Bariloche (CNEA-CONICET), and Universidad Nacional de Cuyo, Avda. E. Bustillo 9500, 8400 - S.C. de Bariloche, Argentina

**ABSTRACT**

Van der Waals heterostructures of 2D layered materials have gained much attention due to their flexible electronic properties, which make them promising candidates for energy, sensing, catalytic, and biomedical applications. Lead iodide (PbI2), a 2D layered semiconductor material belonging to the metal halide family, shows a thickness-dependent band gap with an indirect-to-direct transition above one monolayer. It has emerged as an excellent candidate for photodetectors and is a key component in metal halide perovskites solar cells. In the current work, we investigated the growth dynamics and the real-time correlation between structural and optical properties of $PbI_2$ layers deposited on graphene/SiC(0001) by Molecular Beam Epitaxy. The structural and optical properties are probed respectively by Grazing Incidence Fast Atom Diffraction and Surface Differential Reflectance Spectroscopy. The growth proceeds layer-by-layer in a van der Waals-like epitaxy, with the zigzag direction of $PbI_2$ parallel to the armchair direction of graphene. Both techniques bring evidence of significant modifications of the structural, electronic, and optical properties of the first $PbI_2$ monolayer, characterized by a 1% tensile strain that relaxes over 3 to 5 monolayers. For a single monolayer, Angle-Resolved Photoemission Spectroscopy reveals a charge transfer from graphene to $PbI_2$, demonstrated by an energy shift of the order of 50 meV in the graphene band structure.

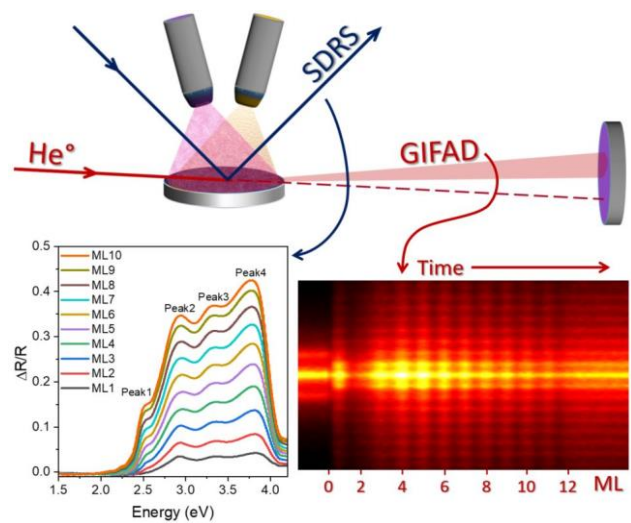

## INTRODUCTION

Since the successful synthesis of graphene and the discovery of its exceptional properties [1,2], two-dimensional semiconductor materials such as transition metal dichalcogenides (TMDs), transition metal carbides (TMCs), and transition metal oxides (TMOs) have attracted enormous interest because of their unique chemical and tunable optoelectronic properties [3–5]. Contrary to graphene, these materials show a sizeable band gap that enables exploration of novel electronic and optical functionalities provided that their interfacial interactions is well understood. In the TMD family, $MX_2$ (with M=W, Mo and X= Se, S) are the most studied despite their low charge-carrier mobility [6,7], which impairs their practical applications in optoelectronics and field-effect transistors. TMD's exhibit a band gap within the range of 1.0–2.0 eV, which restricts their application to photodetection within the wavelength range of 620–1240 nm. To cover the entire detection range of ultraviolet and visible light-emitting diodes, there is a need to explore new types of 2D structures with appropriate band gaps. Lead iodide ($PbI_2$) could be a potential candidate. It has recently attracted much attention due to its important role in hybrid perovskite solar cells [8], photodetectors [9], and non-linear optics [10].

$PbI_2$ is a 2D layered van der Waals (vdW) crystal in its bulk form. Sharing the same three-plane structure of TMDs, a lead plane is sandwiched between two iodine planes, as illustrated in Figure 1. Depending on the stacking configuration, $PbI_2$ has several polytypes in its three-dimensional form, all showing a hexagonal lattice. The well-known and most stable phase at room temperature is the 1T polytype, with lattice parameters $a = b = 4.56$ Å, $c = 6.99$ Å [11]. $PbI_2$ presents a large bulk bandgap of 2.26 –2.4 eV. Theoretical calculations predict that $PbI_2$ has a direct band gap of 2.26 eV for the bulk, which evolves to an indirect band gap of 2.63 eV as the thickness decreases to one monolayer [12,13]. Various forms of $PbI_2$, such as flakes [14], nanoclusters [15], nanoparticles [16], nanodisks [17], and multiwalled nanotubes [18], including thin films [19] have been studied. Many approaches, such as liquid phase exfoliation [17], atomic layer deposition [20], spin coating [21], and Molecular Beam Epitaxy [22] (MBE), are used to elaborate PbI$_2$ thin films. However, MBE allows deposition

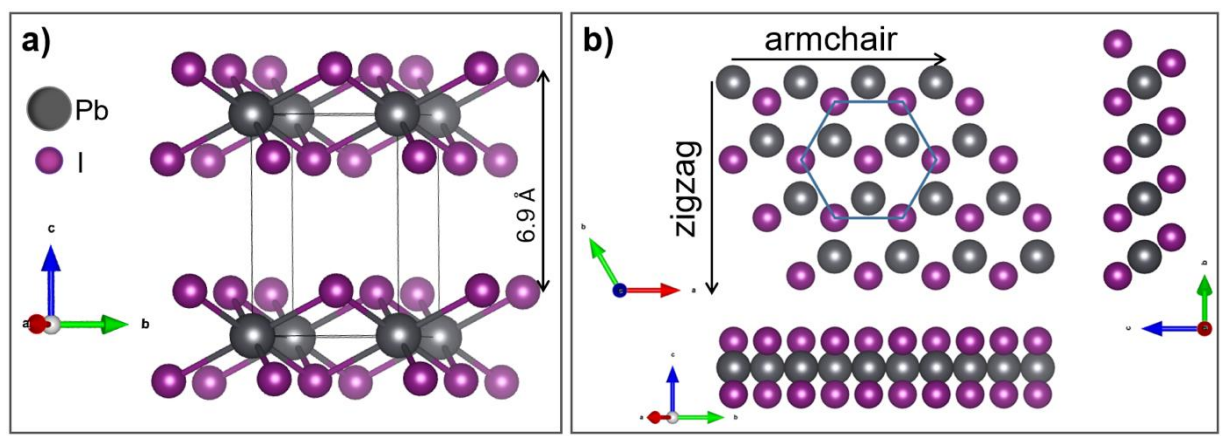

*Figure 1. Atomic structure of PbI2. a) layered structure with the hexagonal unit cell. b) Top view of a monolayer without the bottom iodine plane, and the projections along the armchair and zigzag directions.*

over large surfaces and offers the best control of the layers properties.

In this letter, few lead iodide monolayers were grown on a bilayer graphene/SiC(0001) by MBE in an ultra-high vacuum chamber. To achieve the best control over the growth process, the growing layers are characterized in real-time by synchronizing Grazing Incidence Fast Atom Diffraction (GIFAD) and Surface Differential Reflectance Spectroscopy (SDRS). GIFAD is based on the scattering of He atoms with energy in the range of 0.2–5 keV at typical angles close to 1° [23]. Unlike X-ray or electron diffraction, the very soft nature of the He-surface interaction makes GIFAD exclusively sensitive to the last atomic plane only. It provides information on the surface electronic densities at distances of 2-4 Å with respect to the last atomic planes [24]. These distances are similar to those probed by Scanning Tunnelling Microscopy or Atomic Force Microscopy. Additional information on the principle of GIFAD can be found in the Supporting Information (Figures S2 and S3). SDRS yields information on the relative change in the reflectance of the surface [25], which can then be correlated to the thickness and structural changes of the growing layer provided by GIFAD. Additional characterization techniques include Low Energy Electron Diffraction (LEED) and Angle-Resolved Photoemission Spectroscopy (ARPES).

The GIFAD data reveal a $PbI_2$ growth according to the layer-by-layer mode, with a well-defined alignment of the $PbI_2$ lattice with respect to that of the substrate, resulting in highly crystalline layers. Information derived from both GIFAD and SDRS points to a strong modification of the structural and optical properties of the first $PbI_2$ layer. ARPES on a $PbI_2$ monolayer confirms a peculiar interaction at the interface, characterized by a charge transfer from graphene to $PbI_2$. These findings highlight the strength of the interaction at the $PbI_2$-graphene interface, which far exceeds that expected from vdW coupling.

**RESULTS AND DISCUSSION**

**Growth and structural properties**

Ultra-thin layers of $PbI_2$ were deposited on a graphene/SiC(0001) sample maintained at room temperature; the base pressure in the MBE chamber is in the low $10^{-9}$ mbar range. Sublimation of $PbI_2$ powder is achieved within a resistive heating evaporator. Before deposition, the graphene substrate was cleaned by annealing at 520°C, the surface quality is asserted by the intensity and contrast of the diffraction patterns, in addition to a negligible contribution from diffuse scattering. As depicted in Figure 2, two distinct GIFAD patterns, repeating every 60° azimuthal rotation, are identified as arising from the zigzag and armchair directions (Figure 2a). GIFAD being a projection technique, the observed diffraction patterns result from averaging the structural properties along the beam direction. As a very simple rule, the larger the corrugation (be it of electronic or topographic origin), the more diffraction orders are visible (see Figure S3 in the Supporting Information). Along the zigzag direction (Figure 2b), the diffraction pattern, with a reciprocal vector of 2.95 Å$^{-1}$, corresponds to the graphene honeycomb lattice. The diffraction pattern along the armchair direction (Figure 2c) shows a reciprocal vector of 0.39 Å$^{-1}$, which matches the value expected from the 13x13 Moiré superstructure [26]. In this direction, the electronic corrugation of the graphene honeycomb lattice averages to zero since no diffraction peaks are visible at multiples of ± 5.11 Å$^{-1}$. These results perfectly reproduce those of an earlier GIFAD study on the structure of graphene/SiC(0001) [27]. To resolve the Moiré diffraction pattern, the beam energy was reduced to 350 eV. Figure 2d shows the LEED pattern on the clean graphene sample. Besides the bilayer graphene lattice spots, marked by red circles, the Moiré superstructure is visible through the satellite contributions surrounding the graphene spots [28]. Additional peaks can be attributed to the SiC(0001) substrate [26,27,28].

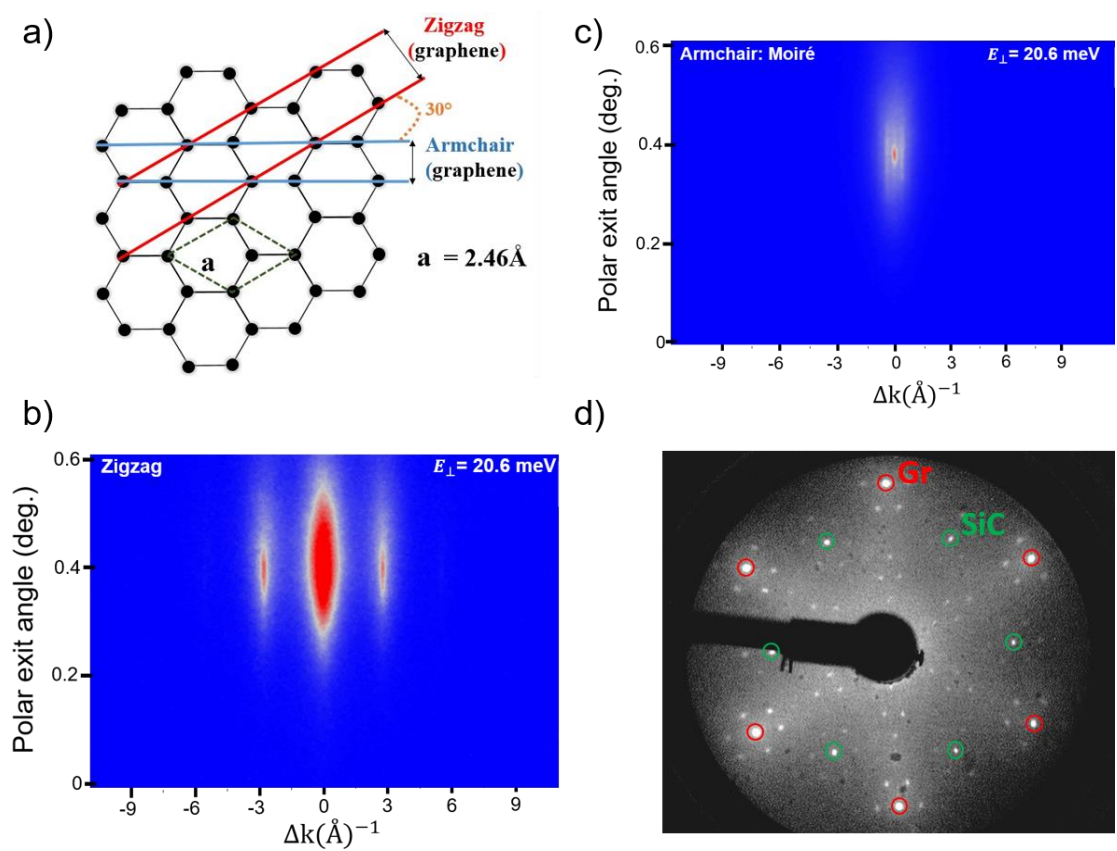

*Figure 2. Atomic structure of graphene/SiC(0001). (a) single layer of graphene with its two crystallographic directions: armchair and zigzag. (b) and (c) are respectively the armchair (moiré) and the zigzag diffraction patterns seen by GIFAD with a 350 eV He beam, (d) LEED pattern of epitaxial graphene at 120 eV; the hexagonal graphene spots are circled in red and those from SiC in green.*

Figure 3a shows the evolution of the GIFAD pattern during deposition of 12 PbI$_2$ monolayers (ML); the probe beam, at 600 eV energy, is aligned along the zigzag direction of graphene. The appearance of a symmetric, with respect to Δk = 0 Å$^{-1}$, and well resolved diffraction pattern from the PbI$_2$ overlayer demonstrates a good alignment between the PbI$_2$ and graphene lattices, with the armchair (PbI$_2$) // zigzag (graphene), as sketched in Figure 3d. This relative alignment is in agreement with observations made by Sinha et al [17] by drop casting PbI$_2$ flakes on suspended graphene. The latter study also mentions the rare occurrence of the opposite alignment, zigzag (PbI$_2$) // zigzag (graphene), which can be completely excluded in our case.

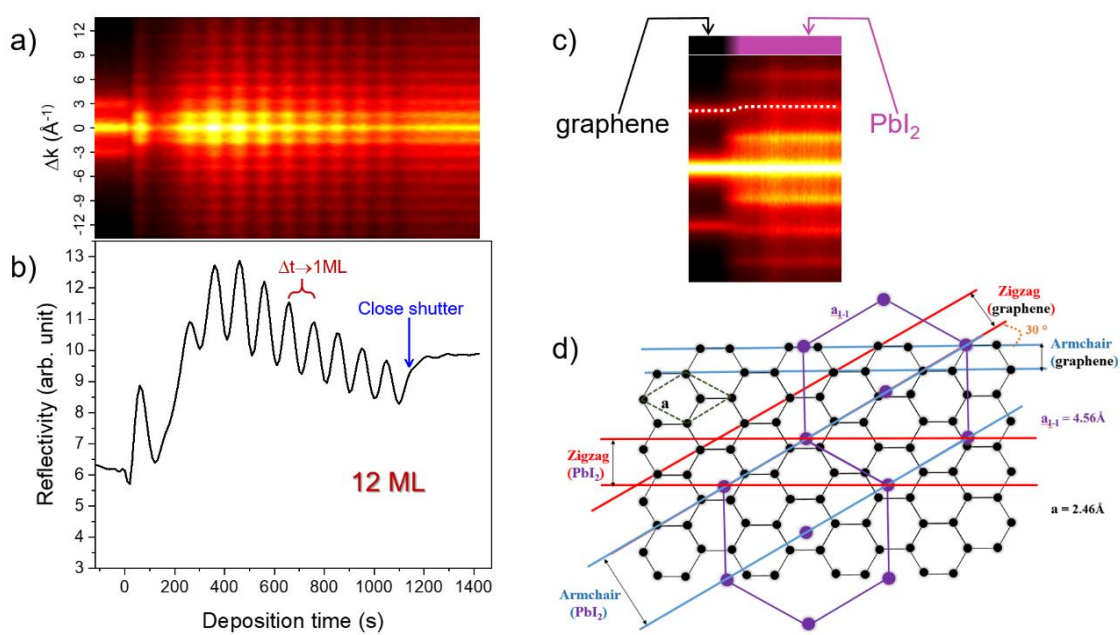

*Figure 3. Evolution with deposition time of a) the diffraction pattern, b) the surface reflectivity during the growth of a 12 ML thick PbI$_2$ layer, the shutter is opened at t=0. b) Highlights the lattice mismatch between the graphene substrate and PbI$_2$, the data is taken from a similar growth sequence; the intensity was normalize to keep constant the 0$^{th}$ order peak intensity. c) Sketch of the relative orientation of the PbI$_2$ lattice with respect to the graphene substrate; the armchair direction of PbI$_2$ is parallel to the zigzag direction of graphene.*

The surface reflectivity, obtained by projecting the 2D image of Figure 3a on the time axis, is shown in Figure 3b. The observed periodic oscillations indicate a layer-by-layer growth mode, and each maximum corresponds to a ML completion. This behaviour is similar to that observed in Reflection High Energy Electron Diffraction (RHEED) in MBE [31], except that GIFAD oscillations are much more robust since their phase does not depend on incidence angle, contrary to RHEED [32]. The measured growth rate is 0.72 ML/min. Figure 3b, extracted from a similar growth sequence, highlights the non-commensurability between PbI$_2$ and graphene, which is also noticeable in Figure 3c. To highlight the relative peak positions at the interface, the intensity was normalized to keep the 0$^{th}$ order peak intensity constant. These observations are characteristic of vdW epitaxy [33] whereby a unique orientation of the overlayer is preserved on the entire substrate despite a large lattice mismatch.

Typical GIFAD images and spectra for the PbI$_2$ multilayers (>5 ML) are shown in Figures 4a and 4b. The spectra result from integrating, along the polar angle, the intensity contained between two Laue circles around the specular angle (0.63°). We notice, as for graphene (Figure 2), that one direction is much more corrugated than the other. However, contrary to graphene, here the armchair direction is the most corrugated and the other direction does show diffraction peaks from the hexagonal PbI$_2$ lattice. Analysis of the diffraction patterns yields reciprocal vectors of 1.59 ±0.01 Å$^{-1}$ and 2.95 ±0.01 Å$^{-1}$ for the armchair and zigzag direction, respectively.

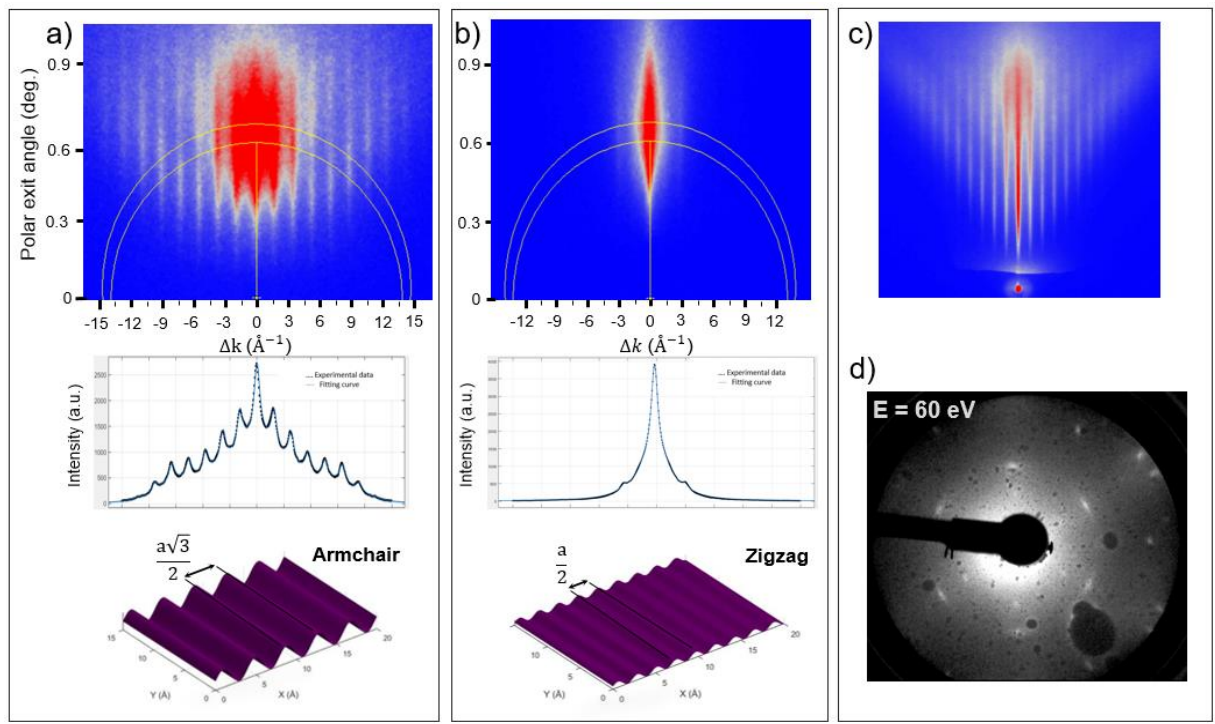

*Figure 4. GIFAD and LEED data from PbI$_2$ layers. a) and b) represent the diffraction patterns and their respective spectrum along the armchair and the zigzag direction respectively. c) diffraction chart obtained by summing up images collected during a polar scan along the armchair direction. d) LEED pattern of one monolayer PbI$_2$ /graphene/SiC(0001).*

We derive a lattice parameter of 4.57 ±0.01 Å, in very good agreement with the reported value of 4.558 Å [34].

As stated above and relying on Figure 3a, the PbI$_2$ layer exhibits a well-defined orientation with respect to the graphene substrate. A closer look at the incidence angle dependence of the diffraction pattern provides a more detailed information. For instance, for the growth of PbI$_2$ on an ultrathin crystalline layer of methylammonium iodide (MAI) deposited on Ag(001), the true epitaxial relationship results in perfectly aligned lattices. Such a unique azimuthal orientation translates into an oscillatory dependence of the diffraction peaks intensity as a function of incidence angle (Figure S4 in the Supporting Information). Although we can distinguish a weak minimum for most diffraction peaks in the diffraction chart of Figure 4c, generated by summing up 50 images acquired at incidence angles in the range 0.08°-1.06°, this is much less marked than in figure S4 of the Supporting Information. We deduce a weak angular dispersion in the relative orientation between the PbI$_2$ layer and the graphene substrate. The LEED pattern of Figure 4d confirms this conclusion; the angular dispersion is estimated to be lower than 5°.

Because of the layered configuration, the structural properties of PbI$_2$ are not expected to vary with thickness [35], we therefore explored whether the substrate had any influence on the structural parameters. According to Figure 5a, the PbI$_2$/graphene interface induces a noticeable tensile strain on the first PbI$_2$ monolayers; this substrate effect stretches the in-plane lattice parameter of the first monolayer by ~1%, with a complete relaxation around 3-5 ML. The observed strain can be explained by the lattice mismatch visible in the inset of Figure 3b, with $(2.g_{PbI2} - g_{Gr})/2.g_{PbI2} = 7.2\%$, combined to the very flexible nature of PbI$_2$ [12]; $g_{PbI2} = 1.59$ Å$^{-1}$ and $g_{Gr} = 2.95$ Å$^{-1}$ are the respective reciprocal vectors. Additional and valuable information can be extracted from the thickness dependence of the diffraction patterns, as those shown in Figures 4a and 4b. In GIFAD, the equipotential energy surface of the He-surface interaction solely determines the relative intensities on the diffraction orders (Figures S2 and S3 of the Supporting Information). In good approximation and thanks to the inertness of He, for incident normal energies larger than 20 meV, this equipotential energy surface follows the electron density profile, an intrinsic property of the surface that contains both atom positions and the corresponding wavefunctions. The relative intensity of the diffraction peaks can thus be considered as a very sensitive fingerprint of the electron density distribution within the crystal lattice. Figure 5b shows these relative intensities as a function of the $PbI_2$ layer thickness. Although the lattice parameter appears to stabilize already near 3 to 5 ML, the electron density distribution continues to evolve until 9 to 10 ML. We notice the strongest perturbation for the first monolayer, visible through the behaviour of the $0^{th}$ and $2^{nd}$ orders. We conclude that the interface interaction is either much stronger than pure vdW and involves chemical effects or, most probably, retains its vdW character but involves a charge transfer allowed by a favourable band alignment [36].

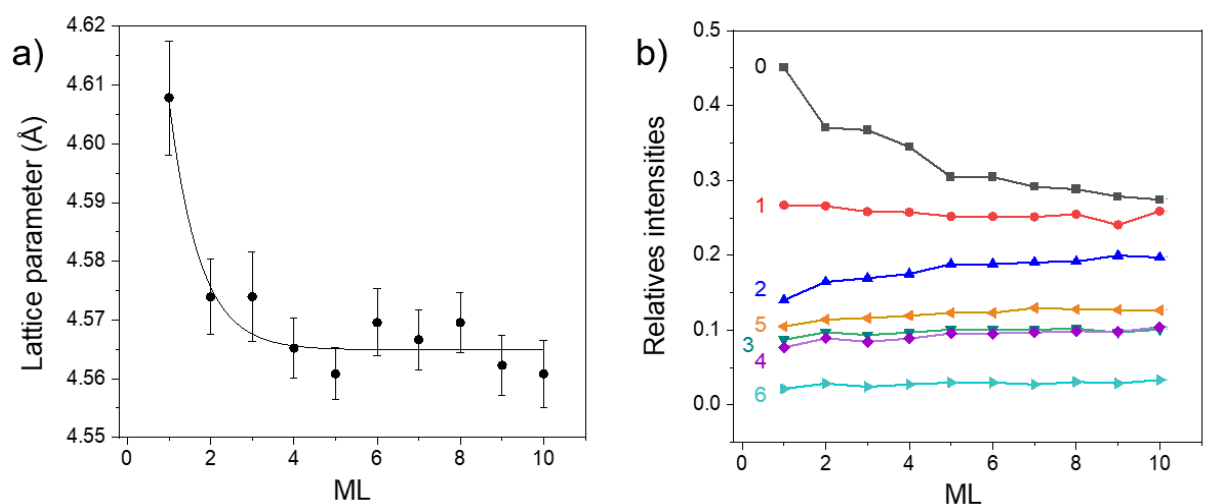

*Figure 5. Dependence of structural properties with layer thickness. a) Lattice parameter, the line is an exponential fit to guide the eye. b) Relative intensities of the diffraction orders.*

**Optical response**

Similarly to the structural properties described in Figure 5 and thanks to the correlated GIFAD and SDRS measurements, we can analyze the thickness dependence of optical properties of the growing layer. As for other 2D materials such as TMD's, isolated $PbI_2$ exhibits a band gap value that evolves with layer thickness. The band gap of $PbI_2$ shows an indirect character for 1 ML and, according to calculations, becomes direct from 2 ML and above [13]. This indirect-to-direct transition is ascribed to the orbital hybridization of iodine atoms from neighbouring layers [17,22], while the change of value is attributed to quantum confinement effects. It remains to be understood how the interface interaction, which gives rise to the modified structural properties identified in Figure 5, further influences the band gap and, more generally, the optical properties.

Figure 6a presents the thickness dependence of the differential reflectance UV/Vis spectrum. Note that the intensity drop observed above 3.93 eV is due the transmission cut-off of the viewport at the air-vacuum interface. These SDRS spectra are directly related to the optical absorption of $PbI_2$ layers thanks to the optical transparency of graphene [37–39]. For 10 ML, we identify four peaks in the spectrum, labelled from 1 to 4, at respective energies of 2.51, 2.93, 3.30, and 3.76 eV. All these peaks were observed in previous studies as characteristic in both absorption and reflectivity of $PbI_2$ [40,41]. Notably, Gahwhiller et al. [42] reported on the reflectance and electroreflectance of $PbI_2$ single crystals at temperatures of 4.5 K and 77 K. They observed peaks at 2.50, 3.09, 3.31, and 3.96 eV and all but the one at 3.09 eV, showed a pronounced narrowing at the lower temperature and was interpreted as due to their excitonic nature. Numerous studies have reported similar findings, and the authors interpreted these excitonic peaks as localized excitations of iodine [42–44]. The peak at 3.09 eV, observed in the photoconductivity of $PbI_2$ by Dugan and Henisch [43] at 3.02 eV is interpreted as an interband transition [44,45]. The latter closely resembles our peak 2 observed at 2.93 eV. In Figure 6b, we display the absorption coefficient, as a function of energy, of each monolayer. A cumulative offset of $1.0 \times 10^5$ is added to all curves above 1 ML. We observe for 10 ML an absorption coefficient of the order of $1.5 \times 10^6$ cm$^{-1}$ at 3.8 eV, which is nearly twice as high as the values reported in the literature [41]. Contrary to the other peaks, the absorption coefficient of peak 2 (at 2.93 eV) shows a maximum value for 1 ML. Figure 6c quantifies the position of the peaks as a function of thickness. We observe two distinct trends: peaks 1, 3, and 4 shift monotonously to lower energy with increasing film thickness. The peak positions shift up to 20-60 meV between 1 and 10 ML, this shift can be attributed to the quantum confinement effects [40]. Similar behaviour were reported for various TMD's [46]. In contrast, peak 2 displays a non-monotonic behaviour; it first shifts to higher energy, from 2.92 eV (at 1ML) to 2.95 eV (from 4ML), and then shifts back to ~2.93 eV at 10ML. The blue shift of this interband transition for a thickness up to 4ML is not related to thickness [40] and could rather be related to the lattice parameter strain observed by GIFAD, which completely relaxes around 4ML.

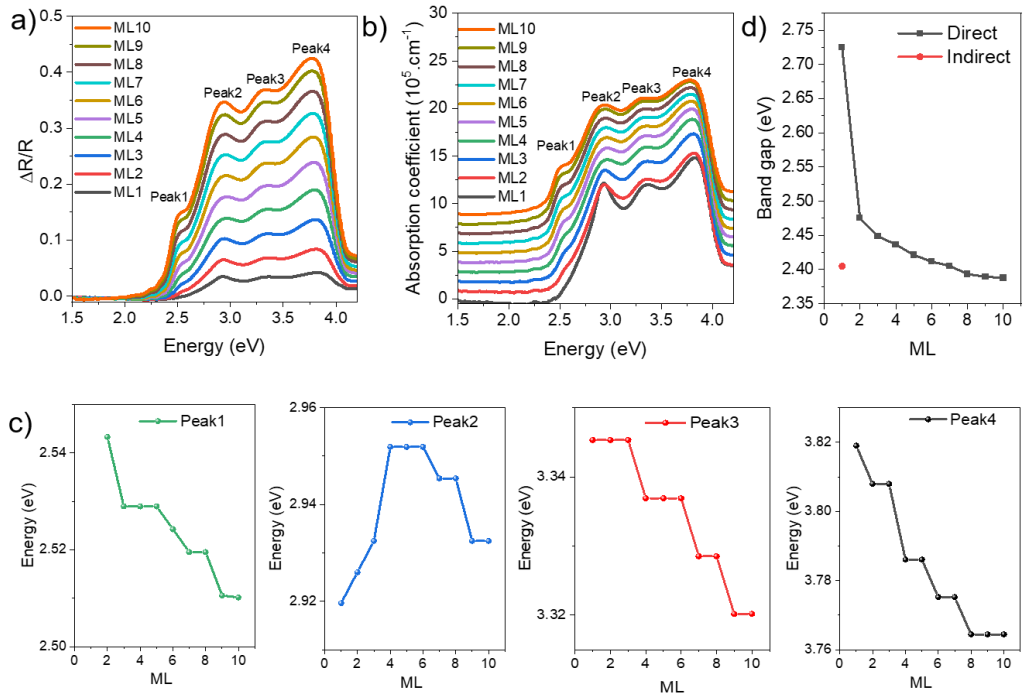

*Figure 6. a) The difference reflectance spectra of PbI$_2$ on graphene during the growth, from 1ML to 10ML. b) Evolution of the absorption coefficient as a function of thickness(for clarity, the absorption spectra were vertically offset by a constant value of 10$^5$ cm$^{-1}$ between successive curves). c) Evolution of peaks position with thickness. d) Dependence of the optical bandgap with thickness.*

Figure 6d shows the variation of the optical band gap, derived from the Tauc plot method [47], as a function of thickness. The results show that the direct band gap decreases rapidly, from a value of 2.72 eV at 1ML to 2.39 eV in the bulk. Assuming an indirect band gap for the first monolayer, the Tauc plot yields a value of 2.40 eV. The observed general trend is similar to the one derived from theoretical calculations of the intrinsic electronic band gap of $PbI_2$ [13,48].

**Photoemission spectroscopy**

$PbI_2$ monolayers and multilayers were further investigated by means of photoemission spectroscopy at the SOLEIL synchrotron facility. We performed core-level photoemission spectroscopy on the 7 ML film to verify its purity and stoichiometry. The energy spectrum is shown in Figure 7. We could associate all main photoemission peaks with the core levels of Pb and I. Most importantly, we note the complete absence of carbon and oxygen, typical contaminants in ambient conditions, thereby attesting the cleanliness of our films and their robustness on transferring using our homemade vacuum suitcase.

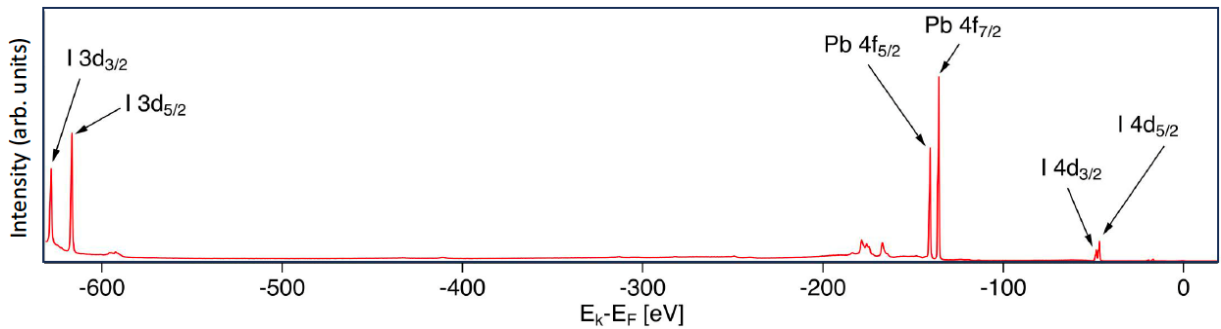

*Figure 7. Core level photoemission spectrum identifying the chemical elements that constitute the $PbI_2$ film.*

In order to clarify the origin of the strongly perturbed electronic density distribution in the first $PbI_2$ monolayer evidenced by GIFAD (Figure 5b), we measured the electronic structure of graphene/SiC(0001) by means of ARPES. We compared the energy-momentum dispersion of pristine graphene and of graphene decorated with 1ML of $PbI_2$. Pristine graphene was obtained after the complete thermal desorption of the $PbI_2$ monolayer upon annealing up to 300°C. Results are shown in Figure 8, where we focused on the electronic structure of graphene at the $\bar{K}$ high-symmetry point of its Brillouin zone, i.e., on the so-called Dirac cone of graphene. The left column refers to the pristine graphene/SiC(0001) substrate, and the right column to graphene covered with 1 ML of $PbI_2$.

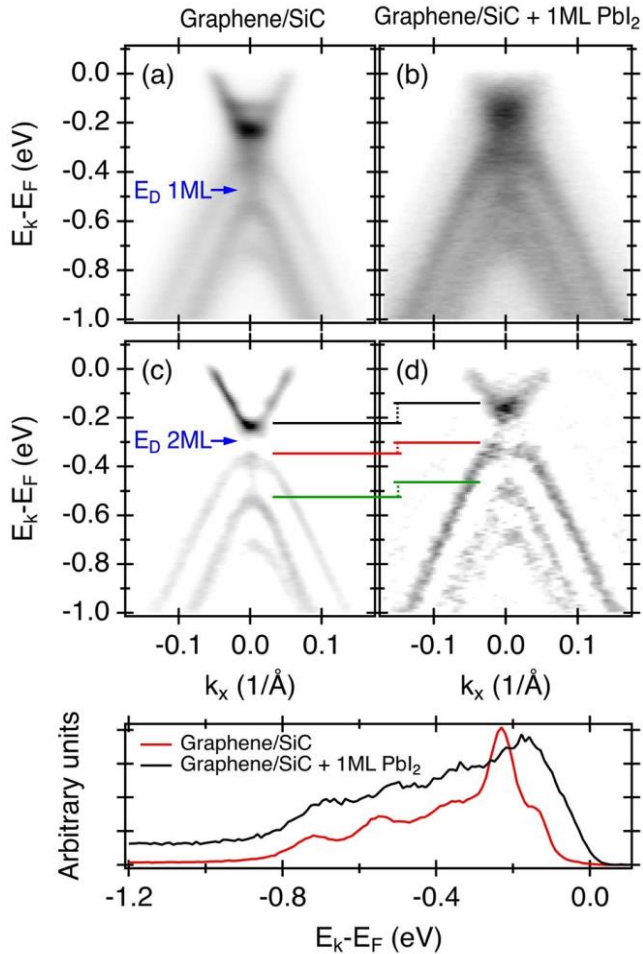

*Figure 8. The energy-momentum dispersion of the bands of graphene near the Fermi level (i.e., the Dirac cone) without [(a), (c)] and with the presence [(b), (d)] of 1 ML of $PbI_2$ on top. The top row is the raw photoemission data, while the middle row is the curvature data (see Methods) in order to enhance the main experimental features. Arrows denote the expected energy position of the Dirac point for 1 ML and 2 ML graphene. Horizontal lines in panels (c) and (d) track the energy shift of the graphene spectral features due to the presence of 1 ML of $PbI_2$. The inset in the bottom row shows an energy dispersion curve (i.e. an intensity profile as a function of energy) acquired at $k_x=0$ Å$^{-1}$ and integrated over 0.01 Å$^{-1}$. One can observe the aforementioned systematic energy shift of the spectral features by comparing the energy positions of the photoemission peaks with (black curve) and without (red curve) the presence of $PbI_2$.*

The electronic structure of the graphene/SiC(0001) shows a multitude of bands, confirming that we do have more than one monolayer of graphene. Although there is a faint linear dispersion with a Dirac crossing point at an approximate binding energy of 500 meV (marked with an arrow in panel a), the dominant spectral signature consists of gapped Dirac cones. The 2D curvature of the photoemission data (lower panels) reveals an energy gap at an approximate binding energy of 300 meV (marked with an arrow in panel c), in excellent agreement with past literature data on bilayer (BL) graphene on SiC(0001) [30,49]. We note that the n-type doping of graphene on SiC is due to charge transfer from the substrate to the graphene layers [50,51] while the gap formation results from inequivalent on-site Coulomb potentials in each layer [52,53]. Despite the good agreement of our data with BL graphene, regarding the gap size and the energy position of the Dirac point, a comparison to the tight-binding bands of the BL reveals that there are supplementary experimental features. Most notably, these are the parabola with a band maximum at binding energy around 550 meV and the inner structure of the electron-like state [51,52,54]. Under our ARPES beam spot, there is therefore an admixture of a third graphene layer. A third layer results in extra bands that can be traced back to the coexistence of rhombohedral (i.e, ABC) and Bernal (i.e., ABA) stacking [52,54]. To sum up, our ARPES data suggest that our substrate is a dominant graphene BL with admixtures of regions with 1 and 3 layers. An admixture of different layers during the epitaxial growth of graphene on SiC is not uncommon and can be tuned with the annealing temperature during the graphene growth [52,55].

In order to evaluate the effect of the first layer of $PbI_2$ on the electronic structure of graphene, we tracked the energy position of various features with and without its presence. As marked by the horizontal lines in panels (c) and (d) of Fig. 8, there is a clear energy shift of all spectral features towards the Fermi level when graphene is decorated with 1 layer of $PbI_2$. The energy shift is corroborated by a comparison of the energy dispersion curves shown in the bottom row of Fig. 8. The red curve is an intensity profile as a function of energy acquired at the $\overline{K}$ high-symmetry point of graphene (i.e. at $k_x = 0$ Å$^{-1}$) when no $PbI_2$ overlayer is present. The black curve is an identical intensity profile acquired on the graphene decorated with 1ML of $PbI_2$. Similar to what was already discussed in panels (c) and (d), a comparison of the energy curves reveals a systematic energy shift of all graphene features towards the Fermi level due to the presence of $PbI_2$. The estimated energy shift is around 50 meV and points towards a charge transfer from the graphene substrate to the first layer of $PbI_2$. We note that changes in the electronic structure of graphene/SiC(0001) were reported in previous literature after the deposition of thin $MoSe_2$ layers [56]. That study was performed on single-layer graphene substrates, with ungapped cones, hence not all details can be directly compared to our work. Nevertheless, as in our work, the charge transfer at the graphene/overlayer interface was reflected into a change in the energy position of the Fermi level and the authors attributed it to the proximity effect through overlapping orbitals in the vdW gap.

In summary, using MBE in combination with real-time correlated GIFAD and SDRS, we have successfully synthesized and characterized in details the growth and opto-structural properties of ultra-thin films of $PbI_2$ on bilayer graphene/SiC from 1 to 10 ML. The GIFAD data reveal a high-quality layer-by-layer growth by a pseudo-vdW epitaxy, with the crystallographic directions of the $PbI_2$ lattice aligned with those of graphene. An in-plane lattice stretching of 1% for the first ML, which relaxes at 4 ML, significantly influences the interband transition of $PbI_2$ at 422 nm. GIFAD also reveals a strongly perturbed electronic density distribution within the first ML, which relaxes slowly around 9-10 ML. These observations suggest a strong interaction at the $PbI_2$/graphene interface. Additional analysis by ARPES demonstrates that this interaction leads to a charge transfer from graphene to $PbI_2$, resulting in a 50 meV energy shift of the graphene band structure. These findings shed a new light on the non-trivial properties of the interface between two vdW materials. This study also provides a benchmark for high quality, large area thin $PbI_2$ layers that could further be used for device development by simply increasing the thickness to the desired value.

**METHODS**

Our experiments are performed in ultra-high vacuum (UHV) at a base pressure in the low $10^{-10}$ mbar range. The samples are mounted on a five-axis manipulator allowing precise azimuthal and polar orientation of the sample with respect to the GIFAD beam.

**Substrate cleaning.** The graphene used in this study is obtained by annealing 4H-SiC(0001) at 1550°C. After the transfer, in UHV is cleaned by performing two annealing cycles of three hours each at 520 °C by means of a tungsten filament heating on the substrate's backside. The temperature is measured with a thermocouple type K connected on the backside of the sample.

**Evaporation.** $PbI_2$ powder (99.999 %, Sigma-Aldrich, used as received) is introduced in a glass crucible, which is radiatively heated inside a three-cell UHV evaporator. The crucible is degassed for hours at 250 °C, close to the evaporation values; a cooling system prevents temperature crosstalk between crucibles.

**GIFAD.** The probe beam is prepared as follows: $He^+$ ions are extracted from an ion source at energies typically in the range of 300–1000 eV. Following neutralization in a He gas cell, a nearly parallel neutral beam is produced by a set of collimating apertures with a size lower than 0.2 mm and directed to the sample surface at grazing angles that can vary between 0.1 and 1.5°. Reflected atoms are collected on a position-sensitive detector made of a set of two Microchannel plates and a phosphor screen, images are captured by a CCD camera.

**SDRS.** The SDRS setup is composed of a Deuterium/halogen lamp UV/Vis light source and two fiber-grating spectrometers (Maya, Ocean Optics, USA). The latter covers a spectral range from 200-1080 nm (1.15- 6 eV) with a resolution of 1 nm. The beam coming from the light source is split into two parts; the first part is called 'Reference', which allows for the correction of lamp spectral drift. The second beam is collimated onto the sample at a typical incidence angle of 45°. The reflected beam is injected into the spectrometer using a 0.22 numerical aperture optical fibre. Incident and reflected beams cross a borosilicate glass viewport at the air-vacuum interface.

The experimental scheme is provided in Figure S1 of the Supporting Information.

**Photoemission.** Angle-resolved and angle-integrated photoemission were conducted at the CASSIOPEE beamline of Synchrotron SOLEIL. The samples were transported in a UHV suitcase where the pressure never exceeded $5 \times 10^{-8}$ mbar. During the photoemission measurements, the pressure stayed below $3 \times 10^{-10}$ mbar, and the temperature was kept at 16 K. We used photon energies of 80 eV and 700 eV for the angle-resolved and core-level photoemission measurements, respectively. In the middle row of Fig. 8, we employed the 2D curvature method [55] to enhance the intensity of broad/weak spectral features. To this end, boxcar smoothing was applied to the raw data using a kernel of 70 meV x 0.012 Å$^{-1}$. The 2D curvature free parameter was set to 0.5

**Supporting Information**

- Principle of GIFAD: experimental scheme, equipotential energy profiles, relationship between diffraction pattern and surface corrugation
- PbI$_2$ layers grown on MAI/Ag(001): Polar scan (diffraction chart)

**Corresponding Authors:**

Christian Perest Sonny Tsotezem : christian-perest.sonny-tsotezem@universite-paris-saclay.fr

Hocine Khemliche : hocine.khemliche@universite-paris-saclay.fr

**ACKNOWLEDGMENT**

This work received funding from Investissements d'Avenir, LabEx PALM (ANR-10-LABX-0039-PALM) and from the Graduate School of Physics (GS-PHOM), Paris-Saclay University.

# Supporting Information

# Correlated Structural and Optical Characterization during Van der Waals Epitaxy of PbI$_2$ on Graphene

C.P. Sonny Tsotezem[1], E. M. Staicu Casagrande[1], A. Momeni[1,2], A. Ouvrard[1], A. Ouerghi[3], M. Rosmus[1], A. Antezak[1], F. Fortuna[1], A. F. Santander-Syro[1], E. Frantzeskakis[1], A.M. Lucero Manzano[4], E.D. Cantero[4], E.A. Sánchez[4], H. Khemliche[1]

[1] Institut des Sciences moleculaires d'Orsay, Université Paris-Saclay, CNRS, 91405, Orsay, France

[2] CY Cergy Paris Université, 95000 Cergy, France

[3] Centre de Nanosciences et de Nanotechnologies, 91120, PalaiseauISMO, Université Paris-Saclay, CNRS, 91405, Orsay, France

[4] Instituto de Nanociencia y Nanotecnología - Nodo Bariloche (CNEA-CONICET), and Universidad Nacional de Cuyo, Avda. E. Bustillo 9500, 8400 - S.C. de Bariloche, Argentina

**Principle of GIFAD.**

The experimental scheme, shown in Figure S1, describes the geometries of GIFAD and SDRS. Both techniques operates simultaneously and are time-synchronized to allow for a reliable correlation between structural and optical properties.

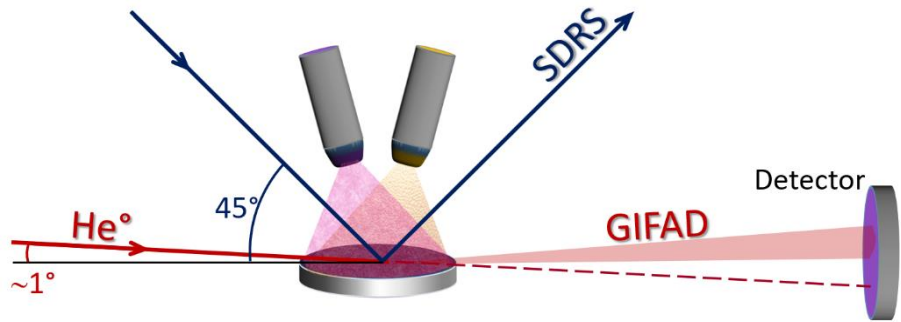

*Figure S1. Experimental scheme combining GIFAD and SDRS.*

Pristine crystalline surface exhibits a two-dimensional corrugation of its electron density. In the grazing geometry used in GIFAD, the corrugation is averaged along the beam direction (x), so only remains the corrugation in the direction perpendicular to the beam direction (y). The latter fully governs the diffraction pattern. More rigorously, the corrugation function measured by GIFAD corresponds to that of the equipotential energy surface Z(y) determined by $(y, Z)) = E_n$, with $E_n = E_0 \sin^2(\theta)$, where $E_0$ is the total incident energy and $\theta$ the incidence angle with respect to the surface plane (see Figure S2). Thanks to the inertness of He, the attractive contribution to the potential can be neglected for normal energies typically greater than 30 meV. The equipotential energy surface therefore follows the isoelectronic density profile [1]. The normal energies range from 1.5 meV [2] to more than 500 meV [3] and correspond to electron densities, around $10^{-3}$ electron/Å$^3$, found at few Å above the last atomic plane.

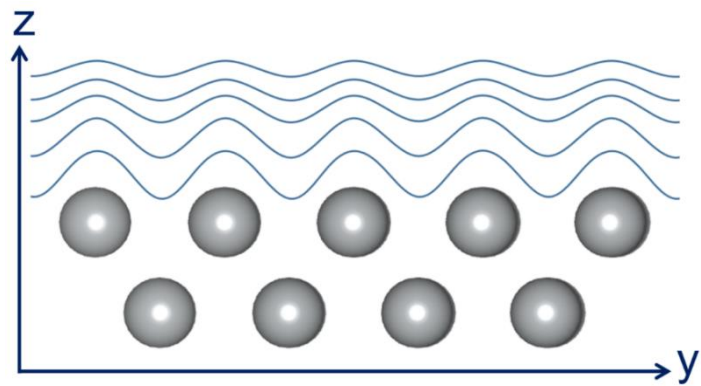

*Figure S2. Equipotential energy lines probed by GIFAD and averaged along the beam direction x. The equipotential energy decreases with Z.*

As depicted in Figure S3, the classical scattering of the incident He atom is solely determined by the shape of the corrugation profile. Neglecting the parallel motion, the maximum scattering angle with respect to the surface normal (so-called rainbow angle) derives directly from the steepest slope of the corrugation function. A high corrugation produces a wider scattering distribution, which accommodates more diffraction peaks. For illustration, for corrugation with a sinusoidal profile, the intensity of the diffraction orders is given by a Bessel function such that $I(n) = J_n(4\pi \frac{h}{\lambda_n})$, with $h$ the corrugation height (difference between top and bottom of the trough) and $\lambda_n$ the de Broglie wavelength related to the beam normal energy $E_n$.

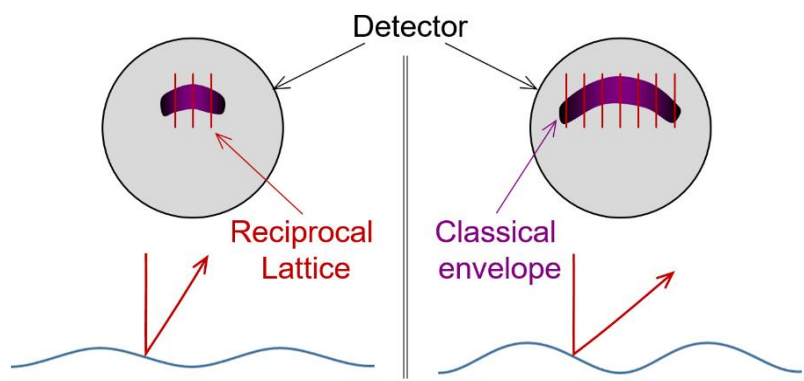

*Figure S3.* Relationship between surface corrugation profile and number of visible diffraction peaks. The banana shaped intensity distribution arises from classically scattering and serves as an envelope of the diffraction pattern.

**PbI2 layers grown on MAI/Ag(001)**

The growth of PbI2 layers on MAI/Ag(001) – MAI: methylammonium iodide, $CH_3NH_3I$ – proceeds layer by layer and results in a unique relative orientation between the MAI and PbI2 lattices. The diffraction chart, obtained by summing up many images acquired at different incidence angles, show clear minima in the intensity of the diffraction peaks (Figure S4). Although scarcely visible, these minima appear to be washed out for the PbI$_2$ layer grown on graphene (Figure 4c in the main text). This is clear signature of a twist mosaicity [4].

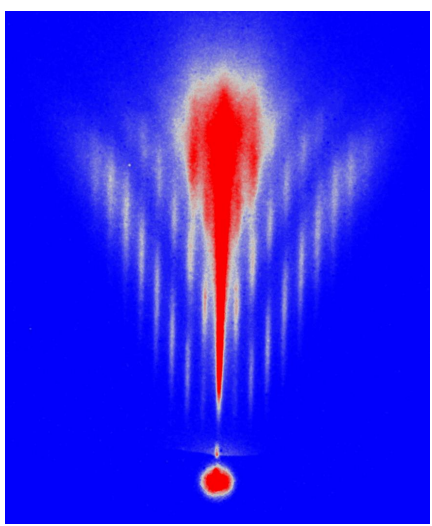

*Figure S4.* Diffraction chart obtained on a PbI$_2$ layer grown on MAI/Ag(001).